\documentclass[%
 aip,
 amsmath,amssymb,
 reprint,%
]{revtex4-1}

\usepackage{amsthm,amsmath,amssymb}
\usepackage{dcolumn, bm, indentfirst, appendix}
\usepackage{graphicx,color,url,orcidlink}
\hypersetup{citecolor=red,colorlinks=true,urlcolor=blue}

\newcommand{\figratio}{0.5}

\begin{document}

\title{Neural Network Solution of Non-Markovian Quantum State Diffusion and
Operator Construction of Quantum Stochastic Process}

\author{Jiaji Zhang\,\orcidlink{0000-0003-2978-274X}}
\affiliation{Zhejiang Laboratory, Hangzhou 311100, China}

\author{Carlos L. Benavides-Riveros\,\orcidlink{0000-0001-6924-727X}}
\email{cl.benavidesriveros@unitn.it}
\affiliation{Pitaevskii BEC Center, CNR-INO and Dipartimento di Fisica, 
Università di Trento, I-38123 Trento, Italy }

\author{Lipeng Chen\,\orcidlink{0009-0002-1541-8912}}
\email{chenlp@zhejianglab.org}
\affiliation{Zhejiang Laboratory, Hangzhou 311100, China}

\begin{abstract}
Non-Markovian quantum state diffusion provides a wavefunction-based framework for modeling open quantum systems. In this work, we introduce a novel machine learning approach based on an operator construction algorithm. This algorithm employs a neural network as a universal generator to reconstruct the stochastic time evolution operator from an ensemble of quantum trajectories. Unlike conventional machine learning methods that merely approximate time-dependent wavefunctions or expectation values, our operator-based approach yields broader applications and enhanced interpretability of the stochastic process. We benchmark the algorithm on the spin-boson model across diverse spectral densities, demonstrating its accuracy. Furthermore, we showcase the operator's utility in calculating absorption spectra and reconstructing reduced density matrices at extended timescales. These results establish a new paradigm for the application of machine learning in quantum dynamics.
\end{abstract}

\maketitle

\section{Introduction}

Multi-dimensional spectroscopy has emerged as a powerful technique for elucidating the structural and dynamical details of chemical reactions. {\cite{mukamel2000arpc, pios2023jpcl, ruetzel2014pnas, kowalewski2017cr, loe2024jcp, panman2017natcomm, bressan2024jpcl, bangert2022natcomm}}
Theoretical descriptions of these spectra require simulating the system's response function under an external laser field. {\cite{mukamel1999, cho2019, oliver2018rsos, gelin2022cr, dorfman2016rmp}} While spectroscopic simulations in gas-phase systems are often performed by directly solving the time-dependent Schr\"{o}dinger equation,
{\cite{pios2024jpcl, gelin2021jctc, sun2020jcp, cho2020jpcb}} the increasing number of degrees of freedom (DOFs) in condensed-phase systems makes a full quantum mechanical treatment prohibitively challenging. In such applications, the theory of open quantum systems provides an efficient strategy. {\cite{weiss2012, vacchini2024}} This framework treats DOFs of lesser theoretical interest as a heat bath, which can be analytically traced out using path integral influence functional theory. The dynamics of the reduced system are then governed by quantum master equation (QME). Well-known examples employing the Markovian approximation for the bath include the Lindblad master equation and the multi-level Redfield equation. {\cite{becker2022prl, kuhn2003jcp, wang2023jctc}} Non-Markovian effects, when necessary, can be incorporated via numerically rigorous approaches like the hierarchical equations of motion (HEOM). {\cite{tanimura2020jcp, yan2014jcp}} These advanced methods have become essential tools for characterizing quantum dynamics in spectroscopic simulations. {\cite{zhang2021jcp, takahashi2023jcp, ikeda2019jcp, duan2022pnas, hein2012jcp, polley2020jpcb}}

Despite their success, a central difficulty with various QMEs is their density-matrix-based formulation, which scales quadratically with system size, making it computationally prohibitive for large systems. To alleviate this limitation, an alternative wavefunction-based approach represents dynamics of open quantum systems by modeling bath effects as quantum stochastic processes. {\cite{weiss2012, milz2021prxq}} In the Markovian case, this leads to the quantum state diffusion (QSD) equation, which can be derived from the stochastic unraveling of the corresponding master equation. {\cite{gisin1992jpa, christie2022jpa, para2024jctc}} This approach has been generalized to the non-Markovian case, the non-Markovian QSD (NMQSD) equation, by incorporating an additional memory integral characterizing the bath effects. {\cite{disi1998pra, strunz1999prl, link2023njp}} However, this integral involves the functional derivative $\delta |\psi(t, {\bm{z}})\rangle / \delta {\bm{z}}$, where $|\psi(t, {\bm{z}})\rangle$ and $\bm{z}$ denote the time-dependent wavefunction and the stochastic noise, respectively. Direct evaluation of this term is only feasible in specific cases, such as harmonic oscillators or via perturbative expansions. {\cite{zhou2024jcp, yu1999pra, zhong2013jcp}} Recently, the hierarchy of pure states (HOPS) method provided an indirect solution by encoding the memory integral into a set of auxiliary equations solved similarly to the HEOM. {\cite{suess2014prl, yu2021natcomm, hartmann2021jpca, gao2022pra, citty2024jcp}} Consequently, while HOPS avoids direct computation of the functional derivative, it inherits HEOM's computational scaling challenges, due to the growth in hierarchical elements. 

In recent years, advances in artificial intelligence (AI) have opened new avenues for applying AI methods, particularly deep neural networks, to scientific research. {\cite{levine2024pnas, krenn2022nrr, dral2024cc, keith2021cr, baum2021jcim}} Within the field of open quantum systems, neural network applications primarily fall into two categories. The first category focuses on function learning, where neural networks approximate time-dependent functions such as density matrix elements and expectation values. {\cite{luis2021jpcl, ullah2022natcomm, mazza2021prr, vicentini2019prl, luchnikov2020prl}} By incorporating quantum mechanics rules, these networks can be trained using short-time input data and subsequently predict long-time dynamics beyond the training time window. To reduce data preparation costs, completely physics-driven and data-free models have been developed by leveraging the time-dependent variational principle. {\cite{norambuena2024prl, rodriguez2024jcp, ullah2022jpcl}} While these approaches have been extensively applied to population and coherence dynamics under time-independent Hamiltonians, their exploration for time-dependent systems with external fields remains limited. The second category involves functional learning, where neural networks approximate functionals describing linear or nonlinear mappings between functions under different conditions. {\cite{chen1995ieee, kovachki2021fno}} A prominent example is the surrogate solver for dynamical equation. {\cite{li2021fno, brunton2024ncs, lu2021nmi}} Such solvers directly generate solutions to target equations (e.g., Schr\"{o}dinger or master equations) from input conditions such as the initial state and external field parameters. Functioning as universal solvers, they are applicable to diverse computational scenarios, including multi-dimensional spectroscopy. {\cite{zhang2024jpcl, zhang2025prr}}

In this work, we aim to develop a neural network architecture capable of solving NMQSD while directly calculating stochastic functional derivatives. To achieve this, we adopt the surrogate model previously developed for solving driven-dissipative quantum dynamics. {\cite{zhang2025prr}} Before proceeding, however, we first summarize key limitations of our previously developed architecture. In earlier work, the mapping from input to output was solely determined by the neural network's intrinsic structure. This approach is inherently black-box, lacking interpretability, and consequently makes it difficult to enhance the surrogate solver's performance by integrating advantages from other algorithms. Improvements were largely restricted to pre-processing input conditions or post-processing output solutions. Furthermore, the network employed a strictly sequential layer structure with a single pathway from input to output. This design proves inefficient for approximating functionals involving multilevel correlations, where highly interwined local (short-range) and global (long-range) correlations coexist. These shortcomings have hampered the model's further development and practical utility.

To address these limitations, we propose a novel surrogate solver algorithm grounded in operator construction. Operators constitute the fundamental building blocks of quantum mechanics; for the NMQSD formalism, we define a stochastic evolution operator $\hat{U}(t, {\bm{z}})$ such that $|\psi(t, {\bm{z}})\rangle = \hat{U}(t, {\bm{z}}) |\psi(0)\rangle$. This operator fully characterizes the stochastic dynamics, and its time derivative ${\rm{d}} \hat{U}(t, {\bm{z}}) / {\rm{d}}t$ recovers the NMQSD equation. Although $\hat{U}(t, {\bm{z}})$ is well defined in principle, its explicit construction has remained unexplored in prior numerical methods, largely due to complexities arising from the stochastic functional derivative. In this work, we demonstrate that $\hat{U}(t, {\bm{z}})$ can be constructed as the output of a surrogate solver, using the noise trajectory ${\bm{z}}$ and physical conditions as inputs. Crucially, once constructed, $\hat{U}(t, {\bm{z}})$ can be decoupled from the neural network, enabling its efficient reuse in diverse computational tasks.

To enhance numerical performance, we also propose a novel neural network architecture for approximating highly correlated functionals in operator construction. This architecture comprises multiple parallel blocks, each serving as an independent pathway between input and output. Within each block, layers are arranged sequentially, with block depths (i.e., layer counts) forming an ascending series. Here, shallow blocks (fewer layers) model short-range local correlations, while deep blocks (more layers) capture long-range global correlations. This design draws inspiration from the U-Net architecture, {\cite{ronneberger2015, wen2022awr}} originally developed to integrate multi-scale contextual information in image segmentation, but adapts it to functional approximation. Compared to traditional sequential structures, our architecture achieves superior expressibility while requiring fewer parameters.

This paper is organized as follows. Section II provides a brief overview of the NMQSD framework and the neural network model. In Section III, we present results for the spin-boson system and analyze the effectiveness of our model. Conclusions are drawn in Section IV.

\section{Methodology}
\label{sec.theory}

\subsection{Non-Markovian Quantum State Diffusion and Operator Ansatz}

We consider a typical system-bath model. The total Hamiltonian is defined as 
\begin{equation}
\hat{H}_{\rm{tot}} = \hat{H}_{\rm{s}} +  \hat{H}_{\rm{b}}  + \hat{H}_{\rm{int}}, 
\end{equation}
where the first term is the system Hamiltonian, 
\begin{equation}
\hat{H}_{\rm{s}} = \sum_{n} \varepsilon_{n} |n\rangle \langle n| + 
\sum_{n \ne n^{\prime}} \Delta_{n n^{\prime}}|n\rangle\langle n^{\prime}|.
\end{equation}
Here, $\varepsilon_{n}$ is the energy of the $n$-th state, and 
$\Delta_{n n^{\prime}}$ represents the interstate couplings. The second term is the Hamiltonian of a harmonic heat bath, 
\begin{equation}
\hat{H}_{\rm{b}} = \sum_{j}  \frac{\hat{p}_{j}^2 + 
\omega_{j}^2 \hat{x}_{j}^2}{2} ,
\end{equation}
where $\hat{p}_{j}$, $\hat{x}_{j}$, and $\omega_{j}$ are the
momentum, coordinate, and frequency of the $j$-th oscillator, respectively. The final term is the system-bath interaction Hamiltonian, 
\begin{equation}
\hat{H}_{\rm{int}} = -\hat{V} \sum_{j} g_{j} \hat{x}_{j},
\end{equation}
where $\hat{V}$ represents the system-part interaction operator, and 
$g_{j}$ is the coupling constant between the system and the $j$-th oscillator. The effect of the heat bath on the system is fully characterized by the bath correlation function,
\begin{equation}
\alpha(t) = \frac{1}{2 \pi} \int_{0}^{\infty} {\rm{d}} \omega \, J(\omega) \left[ 
\coth\left( \frac{\beta\omega}{2} \right) \cos(\omega t) - i \sin(\omega t) \right],
\end{equation}
where $J(\omega)$ is the spectral density function (SDF), and 
$\beta = 1 / k_{B} T$ is the inverse temperature with $k_{B}$ being the Boltzmann constant. Henceforth, we assume the initial state $\hat{\rho}_{\rm{tot}}(0) = |\psi_0\rangle \langle \psi_0| \otimes \hat{\rho}_{b}^{\rm{eq}}$, 
where $|\psi_0\rangle$ is a chosen state of the system, and
$\hat{\rho}_{b}^{\rm{eq}}$ is the equilibrium distribution of the heat bath.

Next, we introduce a complex Gaussian stochastic process $z = z_t$, with mean and correlations given by  
\begin{equation}
\begin{gathered}
\mathbb{E}\left[ z_t \right] = \mathbb{E}\left[ z_t z_s \right] = 0  \\
\mathbb{E}\left[ z_t z_s^{\ast} \right] = \alpha(t - s) ,
\end{gathered}
\label{eq.noise_nonmarkov}
\end{equation}
where $\mathbb{E}[ \cdots ]$ denotes the ensemble average over the noise $z_t$.
The reduced density operator can be obtained from the ensemble average over trajectories of pure states via {\cite{weiss2012, gisin1992jpa}}
\begin{equation}
\hat{\rho}_{\rm{s}}(t) = {\rm{Tr}}_{\rm{b}}\{ \hat{\rho}_{\rm{tot}}(t) \} =
\mathbb{E}\left[ |\psi(t, z) \rangle \langle \psi(t, z) | \right].
\label{eq.reduced_density}
\end{equation}
The time evolution of these stochastic states is governed by the NMQSD equation {\cite{disi1998pra, strunz1999prl}}
\begin{equation}
\begin{split}
\partial_t | \psi_t \rangle &= -i \hat{H}_{\rm{s}}  | \psi_t\rangle + 
\hat{V}  z_{t}^{\ast} \, | \psi_t \rangle \\
&- \hat{V}^{\dagger} \int_{0}^{t} {\rm{d}}t^{\prime} \, \alpha(t - t^{\prime})
\frac{\delta  | \psi_t \rangle}{\delta z_{t^{\prime}}^{\ast}},
\end{split}
\label{eq.nmqsd_linear}
\end{equation}
where we use the abbreviation 
$|\psi_t \rangle \equiv |\psi(t, z) \rangle$ and the  initial condition 
$| \psi(t = 0, z) \rangle \equiv |\psi_0 \rangle$. The last integral term captures the non-Markovian memory effect. This term vanishes in the Markovian limit where $\alpha(t) \propto \delta(t)$ , causing Eq.{\eqref{eq.nmqsd_linear}} to reduce to the standard stochastic unraveling of a Lindblad master equation. 

Equation {\eqref{eq.nmqsd_linear}} can suffer from numerical instability as the norm  $|| \psi_{z}(t) || $ tends to zero for most trajectories. To ensure numerical stability, we perform a Girsanov transformation to obtain a nonlinear equation for the normalizable state $|\tilde{\psi}_{t} \rangle = | \tilde{\psi}(t, z) \rangle$ 
\begin{equation}
\begin{split}
\partial_t | \tilde{\psi}_t \rangle &= -i \hat{H}_{\rm{s}} | \tilde{\psi}_t \rangle + 
\hat{V} \tilde{z}_{t} \, | \tilde{\psi}_t \rangle \\
&- \left( \hat{V}^{\dagger} - \langle \hat{V}^{\dagger}\rangle_t  \right) \int_{0}^{t} 
{\rm{d}}t^{\prime} \, \alpha(t - t^{\prime}) \frac{\delta  | \tilde{\psi}_t \rangle}
{\delta \tilde{z}_{t^{\prime}}}.
\end{split}
\label{eq.nmqsd_nonlinear}
\end{equation}
Here, the expectation values $\langle\cdot\rangle_t$ are calculated using the normalized state $|\tilde{\psi}_t\rangle/\sqrt{\langle\tilde{\psi}_t|\tilde{\psi}_t\rangle}$, and we have introduced a shifted noise
\begin{equation}
\tilde{z}_{t} = z_{t}^{\ast} + \int_{0}^{t} {\rm{d}}t^{\prime} \, \alpha^{\ast}(t-t^{\prime}) 
\langle \hat{V}^{\dagger}\rangle_{t^{\prime}} .
\label{eq.nmqsd_shift_noise}
\end{equation}
For the remainder of this work, we will always consider this nonlinear equation unless otherwise specified. 

We now introduce an operator ansatz for the time evolution in NMQSD. For simplicity, we consider only a discrete-time version with time stamps $t_{n} = n \delta_t$. The continuous form can be recovered by taking the limit $\delta_t \to 0$. We first define a stochastic time evolution operator for Eq.~{\eqref{eq.nmqsd_nonlinear}},
\begin{equation}
|\tilde{\psi}_{t_n}\rangle = \hat{U}(t_n, {\bm{z}}_{[0:n]}, {\psi}_{0}) |{\psi}_{0}\rangle, 
\label{eq.operator_evolve}
\end{equation}
where the matrix representation for discrete time points is 
${\bm{z}}_{[m:n]} = \{ z_{t_m}, z_{t_{m+1}}, \cdots, z_{t_n} \} $. Note that the additional dependence on the initial state ${\psi}_{0}$ arises from the shifted noise in the nonlinear equation; this dependence vanishes in the linear formulation.

\subsection{Neural Network Model}

\begin{figure}
\centering
\includegraphics[width=\figratio\textwidth]{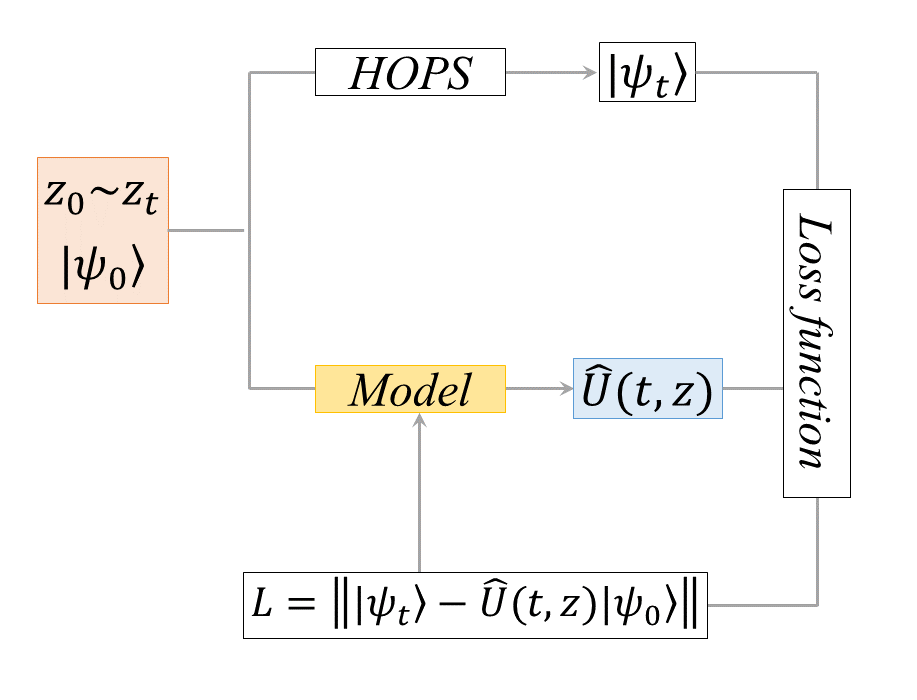}
\caption{Schematic of the operator construction algorithm workflow. Stochastic noise $z_0 \sim z_t$ and the initial quantum state $|\psi_0\rangle$ (red) serve as inputs to a neural network model (yellow). The network's output is the time-evolution operator $\hat{U}(t, z)$ (blue). The training process involves the optimization of a loss function $L$, where the training data comprises quantum states $|\psi_t\rangle$ obtained from numerical integration of the HOPS equation.}
\label{fig.workflow}
\end{figure} 

The time evolution operator can be systematically constructed using a neural network. The primary steps of this operator construction are presented in Fig.~{\ref{fig.workflow}}. At a specific timestamp $t_{n}$, we introduce a matrix representation for the time-dependent wavefunction,  
${\bm{\psi}}_{n} \sim | \tilde{\psi}_{t_n} \rangle$, and for the operator 
${\bm{U}}_{n}\sim \hat{U}(t_n, {\bm{z}}_{[0:n]}, {\psi}_{0})$. We define a neural network such that 
\begin{equation}
{\bm{U}}_{n} = \mathcal{G}_{\theta} \left( {\bm{x}}_{n} \right),
\label{eq.neural_network}
\end{equation}
where $\theta$ represents all learnable parameters, and ${\bm{x}}_{n} = \{ t_n, {\bm{z}}_{[0:n]},  {\bm{\psi}}_{0} \}$ is the collective input. In the numerical implementation, the arguments ${\bm{x}}_{n}$ serve as the network's input, and the corresponding matrix on the left-hand side is the output. The time evolution is then obtained by performing a matrix-vector multiplication with the outputted operators, following Eq.~{\eqref{eq.operator_evolve}}.

Using operators as the neural network output offers several potential advantages. In our previous works {\cite{zhang2024jpcl, zhang2025prr}} and other neural solvers {\cite{li2021fno, lu2021nmi}}, the network directly outputs the solution of the target differential equation. When applied to the NMQSD equation, this direct parametrization scheme requires the time-dependent wavefunction to be the output (e.g., ${\bm{\psi}}_{n} = \mathcal{G}_{\theta}({\bm{x}}_{n})$). Consequently, the operators defined above would correspond to the internal structure of the model itself. Owing to the black-box nature and lack of interpretability of neural networks, further use of these operators for other tasks is nearly impossible. In contrast, the indirect parametrization used in this work (Eq.{\eqref{eq.neural_network}}) employs the operator itself as the output. The solution must be obtained through post-processing; for instance, the time evolution ${\bm{U}}_{n} = \mathcal{G}_{\theta}({\bm{x}}_{n} )$ is computed first and then applied as ${\bm{\psi}}_{n} = {\bm{U}}_{n} {\bm{\psi}}_{0}$. Once constructed, operators like ${\bm{U}}_{n}$ can be decoupled from the neural network, offering significantly greater flexibility for use in other computational tasks.

\begin{figure}
\centering
\includegraphics[width=\figratio\textwidth]{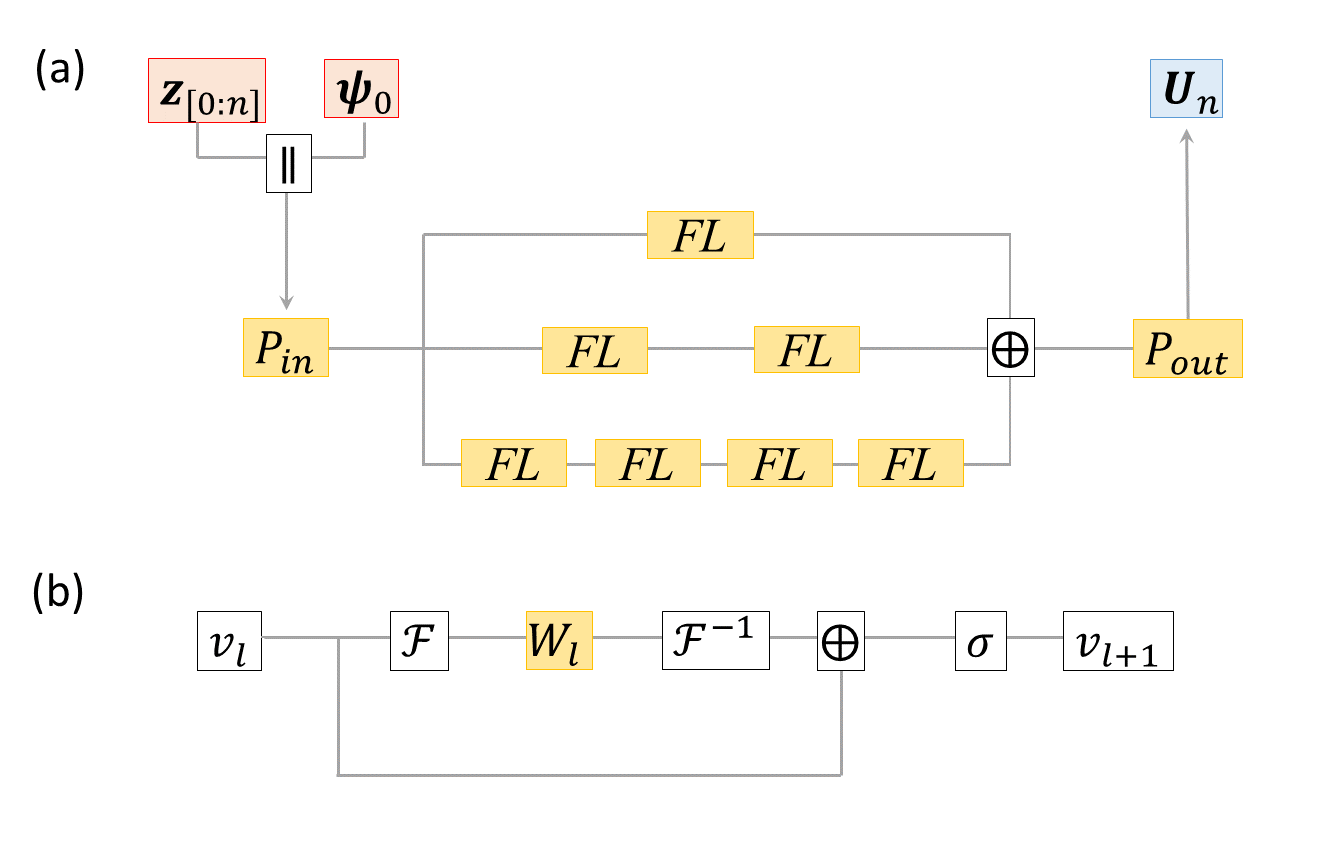}
\caption{(a) Architecture of the neural network model $\mathcal{G}_{\theta}$. Input arguments ${\bm{z}}_{[0:n]}$ and ${\bm{\psi}}_{0}$ are shown in red and are processed through a series of operations to produce the output operator ${\bm{U}}_{n}$ at time $t_{n}$ (shown in blue). The operator `$| |$' denotes vector concatenation, and `$+$' denotes an element-wise sum.  (b) The structure of a single Fourier layer (FL). The operations $\mathcal{F}$ and $\mathcal{F}^{-1}$ denote the Fourier transform and its inverse, respectively, while $\sigma$ is a nonlinear activation function. The  learnable parameters are those in linear projection matrices $P_{\mathrm{in/out}}$ and the point-wise convolution kernel $W_{l}$.}
\label{fig.model}
\end{figure} 

Next, we describe the neural network's architecture, a schematic of which is presented in Fig.~{\ref{fig.model}}. The model's fundamental building block is the Fourier layer (Fig.~{\ref{fig.model}}(b)), which has been widely used for operator learning tasks. This structure performs spectral convolution by applying learnable weights in the Fourier space to capture non-local correlations. To model correlations at different scales, we employ a U-Net structure with several parallel blocks. Each block consists of cascading layers of a specific depth (number of layers). In this configuration, a shallow block with fewer layers captures short-range local correlations, while a deep block with more layers captures long-range global correlations. Compared to a simple linear chain of layers, this mixture of blocks with varying depths mimics a functional series expansion, allowing for a more efficient description of highly mixed correlations.      

To train the neural network, we generate a dataset of time evolution states ${\bm{\psi}}_{n}$ using the HOPS method (see Appendix \ref{sec.app.hops} for details). The loss function for training the time evolution operator is defined as 
\begin{equation}
L^{(1)} = \frac{1}{N_t} \sum_{n} \left| \left| {\bm{\psi}}_{n} - \mathcal{G}_{\theta}
({\bm{x}}_{n}) {\bm{\psi}}_{0} \right|\right|. 
\end{equation}
In contrast to our previous work,{\cite{zhang2024jpcl}} we do not use a physics-informed loss function (PILF), which is data-free and defined solely by the residual of the equation. This is because PILF requires automatic differentiation to compute the Jacobian of the model's output with respect to all inputs ${\bm{z}}_{[0:n]}$. For our network architecture, which involves Fourier transforms, this process is computationally intensive and memory-demanding. Therefore, limited by our computational resources, we rely solely on the precomputed dataset for training in this work.  

\section{Numerical result}
\label{sec.result}

In this section, we test our neural network model on the spin-boson model. The system Hamiltonian is 
\begin{equation}
\hat{H}_{\rm{s}} = \frac{\omega}{2} \hat{\sigma}_{z} + \Delta \hat{\sigma}_{x},
\end{equation}
where $\hat{\sigma}_{x}$ and $\hat{\sigma}_{z}$ are Pauli matrices. We set $\omega = 1$ and $\Delta = 0.5$ for all simulations. The system-part interaction operator is given by $\hat{V} = \hat{\sigma}_{z}$, with the spectral density function and inverse temperature ($\beta$) specified later. We consider several initial states of the form $|\psi_0\rangle = a |1\rangle + b |2\rangle$, where the coefficients $a$ and $b$ are chosen from $\{0, \pm 1, \pm i\}$ and normalized. Non-zero values for both $a$ and $b$ are necessary for studying population dynamics. 

For the numerical implementation, we define a time grid from $t=0$ to $t_{\max}=10$ with a step size of $\delta t=0.01$. The model is trained and evaluated exclusively within this interval. The network architecture consists of three Fourier blocks with an increasing number of layers (1,2,4) and a latent space dimension of 32. Each block performs a one-dimensional Fourier transform on its inputs. To ensure model compactness and prevent overfitting, we truncate the expansion, explicitly including only the 256 lowest-frequency modes with learnable weights. All linear projections are implemented as 4-layer feedforward networks with 128 hidden channels. This configuration results in a total of approximately 3 million trainable parameters. We employ Gaussian Error Linear Units (GELU) as the activation function throughout the network. All models are trained for up to $10^{5}$ epochs using the AdamW optimizer with a learning rate of $10^{-4}$.

\subsection{Operator construction for Drude SDF}
\label{sec.result.drude}

We begin by considering a Drude-type SDF, defined as 
\begin{equation}
J(\omega) = \frac{2 \lambda \gamma \omega}{\omega^2 + \gamma^2},
\end{equation}
where $\lambda$ is the reorganization energy and $\gamma$ is the inverse correlation time. We keep $\lambda = 0.1$ and $\gamma = 1.0$ fixed, representing a case of moderate system-bath coupling. The inverse temperature is set to $\beta = 0.2$, $1$, and $5$, representing high, intermediate, and low-temperature regimes, respectively. For each value of $\beta$, we independently train a neural network model for operator construction using the settings described previously. To generate the training data, we first produce 7000 random noise trajectories, $\{ z_{t} \}$, and then compute their time evolution using the HOPS method. We use 5000 of these samples for training and reserve the remaining 2000 for validation. After training, the model can construct the evolution operator for any given noise trajectory $z_t$, at the corresponding temperature with high accuracy for the chosen initial states.      

To quantify the model's preformance, we define a time-dependent error metric
\begin{equation}
L(t_n) = \mathbb{F}\bigl| 1 - \langle \bm{\psi}_{n}| \hat{U}(t_{n}) |\psi_0\rangle  \bigr|,
\label{eq.time_evolve_error}
\end{equation}
where ${\bm{\psi}}_{n}$ is the reference state computed by HOPS, and $\hat{U}(t_{n})$ is the evolution operator constructed by the model. The functional $\mathbb{F}$ represents an operation performed over the entire validation set. 

\begin{figure}
\centering
\includegraphics[width=\figratio\textwidth]{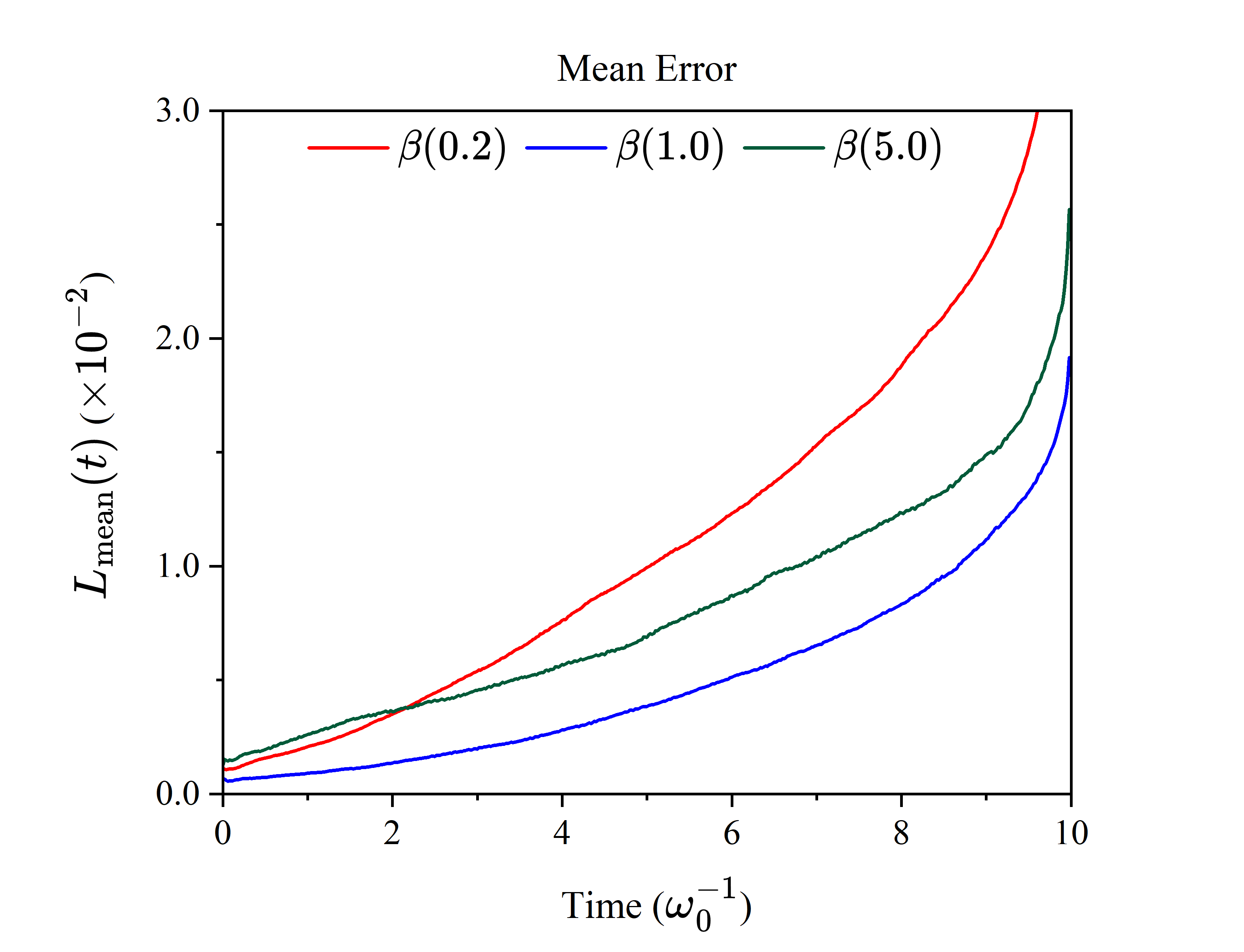}
\caption{Mean error for the  Drude SDF at $\beta = 0.2$ (red), $1.0$ (blue), and $5.0$ (green), respectively.}
\label{fig.mean_drude}
\end{figure} 

We first take $\mathbb{F}$ to be the mean absolute error, which reflects the model's overall accuracy across all validation samples. Figure {\ref{fig.mean_drude}} shows this time-dependent mean error for various $\beta$. A a universal approximator, the trained model achieves a consistently low error at all temperatures. For most of the time interval, $t \in (0, 0.8 t_{\rm{max}})$, the mean error increases nearly linearly with time, a consequence of the growing complexity in the memory integral. However, the error increases more rapidly, almost quadratically, near the boundary $(t > 0.9 t_{\rm{max}})$. This anomaly arises from discontinuities introduced by the truncation of Fourier modes when applying the learnable weights, which affects the fast Fourier transform. While increasing the model size (e.g., using more layers or Fourier modes) can systematically mitigate this numerical issue, our current model already achieves a mean error of approximately 2\%, demonstrating its overall accuracy for unseen noise trajectories at all temperatures.

\begin{figure}
\centering
\includegraphics[width=\figratio\textwidth]{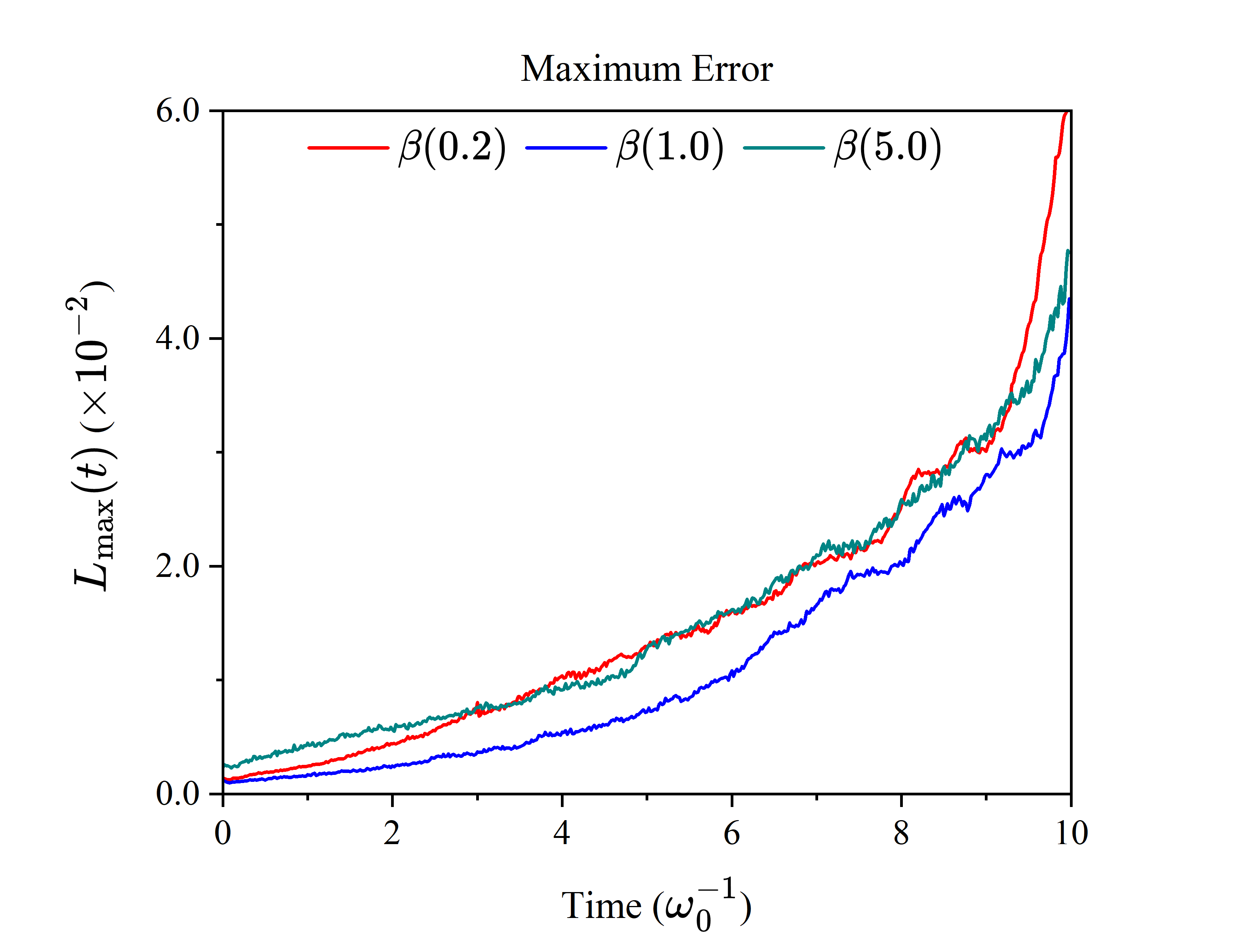}
\caption{Maximum error for the Drude SDF at $\beta = 0.2$ (red), $1.0$ (blue), and $5.0$ (green), respectively.}
\label{fig.max_drude}
\end{figure} 

Next, we consider the case where $\mathbb{F}$ is the maximum operation, which identifies the largest error and thus the worst-case performance on the validation set. As shown in Fig.~{\ref{fig.max_drude}}, the maximum error is nearly twice the mean error, indicating that our model struggles with a subset of extreme noise trajectories. We attribute  this to the inherent structure of our neural network. As outlined in Eq.~{\eqref{eq.noise_nonmarkov}}, the noise $z_t$ is a Gaussian random variable with a time-dependent covariance matrix. Accurately constructing the evolution operator $\hat{U}(t_n, {\bm{z}}_{[0:n]}, \psi_0)$ for any noise trajectory therefore requires an understanding of its probability distribution. However, the cornerstone of our model is a deterministic Fourier layer, which is inherently less efficient at processing probabilistic inputs. The most straightforward way to reduce the number of these extreme outliers is to use a larger training set. A more advanced solution should be to develop a probabilistic neural network architecture specifically designed for the operator learning task.  

\subsection{Operator construction for Brownian SDF}

\begin{figure}
\centering
\includegraphics[width=\figratio\textwidth]{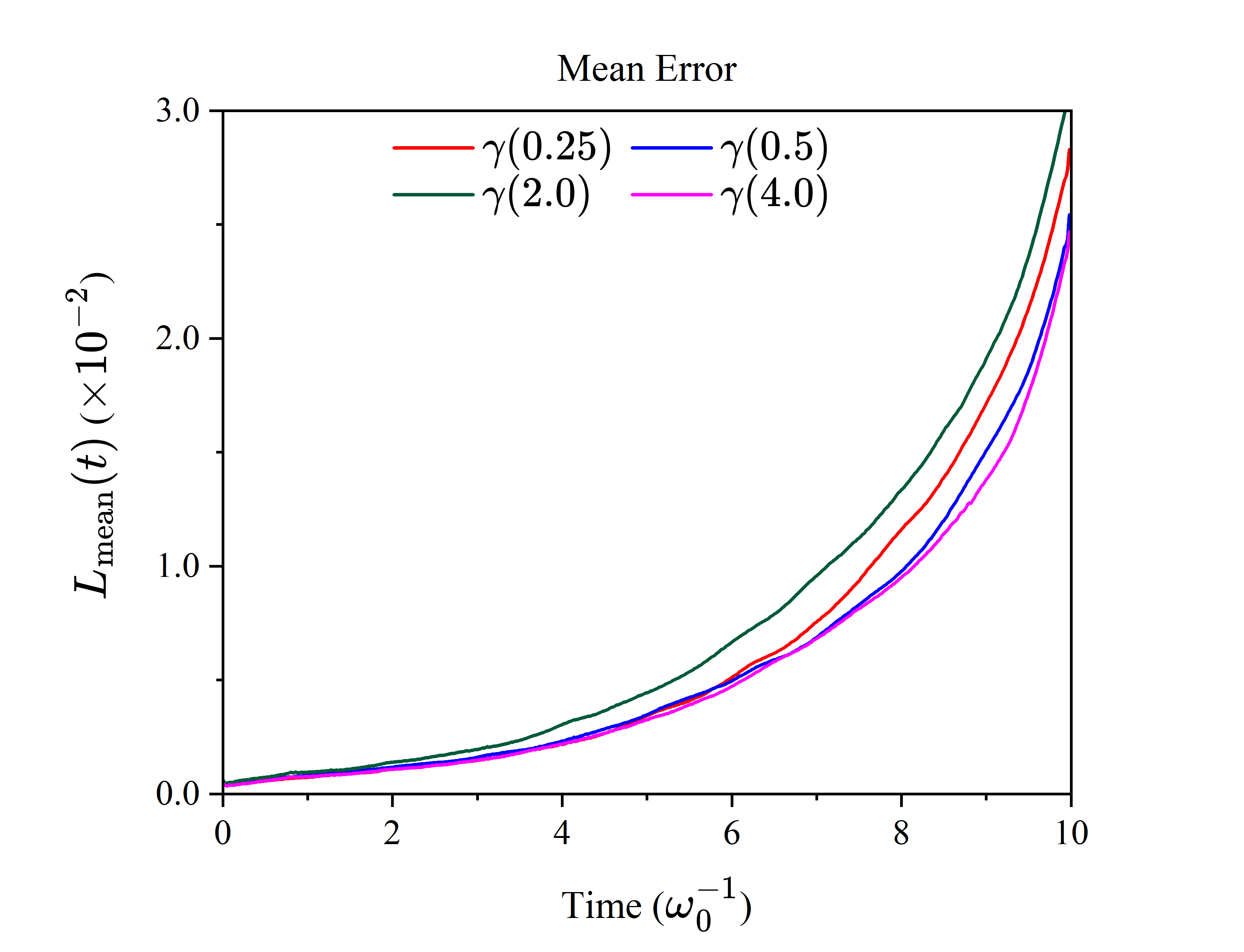}
\caption{Mean error for the Brownian SDF at $\gamma$=0.25 (red), 0.5 (blue), 2.0 (green), 4.0 (magenta), respectively.}
\label{fig.mean_brownian}
\end{figure}

\begin{figure}
\centering
\includegraphics[width=\figratio\textwidth]{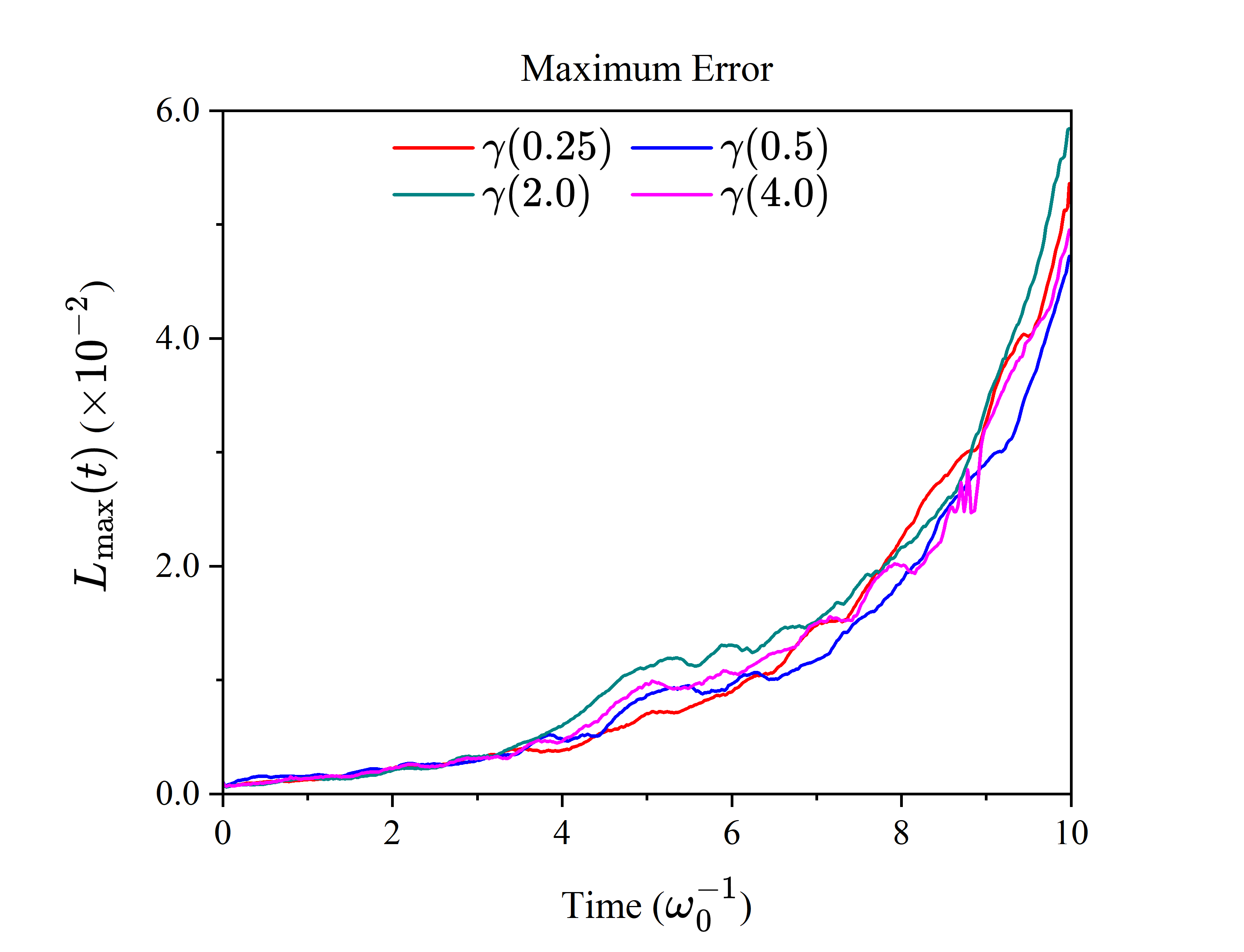}
\caption{Maximum error for the Brownian SDF at $\gamma$=0.25 (red), 0.5 (blue), 2.0 (green), 4.0 (magenta), respectively.
}
\label{fig.max_brownian}
\end{figure}

Next, we consider a Brownian-type SDF, defined by 
\begin{equation}
J(\omega) = \frac{2 \lambda \gamma \omega_b^2 \omega}
{(\omega^2 - \omega_b^2)^2 + \gamma^2\omega^2}.
\end{equation}
This SDF models an effective environment where the system interacts with a primary harmonic oscillator of frequency $\omega_b$, which is itself coupled to a secondary Ohmic heat bath with friction constant $\gamma$. In our simulation, we set the parameters $\lambda = 0.1$, $\omega_b = 0.5$ and $\beta = 1$. We evaluate our model for several underdamped cases $(\gamma = 0.25$ and $0.5)$ and overdamped cases $(\gamma = 2.0$ and $4.0)$. The datasets are generated using the same procedure as in the Drude case, but with this new SDF and its corresponding correlation function $\alpha(t)$. The resulting mean and maximum errors of the trained model are shown in Figs.~{\ref{fig.mean_brownian}} and {\ref{fig.max_brownian}}, respectively. Analogous to the Drude case, our model achieves an overall mean error of $\sim 1.5\%$ and a maximum error of $< 3 \%$ for most of the time interval. This demonstrates the broad applicability of our operator construction algorithm across different types of SDFs.

In all the cases discussed above, a separate model was trained for each specific correlation function $\alpha(t)$. As shown in Eq.~{\eqref{eq.nmqsd_nonlinear}}, the time evolution operator for NMQSD is determined solely by the noise $z_t$, the initial state $\psi_0$, and the correlation function $\alpha(t)$. Since neural networks are universal approximators, it is theoretically possible to train a single, sufficiently large model that is applicable to a broad class of functions $\alpha(t)$, which are uniquely defined by the SDF and inverse temperature $\beta$. However, implementing such a model in practice presents several numerical challenges. A primary difficulty stems from the complex role of $\alpha(t)$: as illustrated in Eq.~{\eqref{eq.nmqsd_nonlinear}}, it not only explicitly governs memory effects through the last term but also uniquely defines the covariance of the noise $z_t$ via Eq.~{\eqref{eq.noise_nonmarkov}}. Furthermore, using the correlated inputs $\alpha(t)$ and $z_t$ simultaneously may lead to overfitting. Consequently, the development of such a comprehensive model remains a key objective for future work.

\subsection{Application of Constructed Operator}

Once constructed, the time evolution operator can be decoupled from the neural network, allowing for convenient use in other tasks. Hereafter, we use the Drude SDF as an example, employing the same dataset and models as in Section {\ref{sec.result.drude}}. We briefly illustrate two typical applications: (1) the calculation of linear absorption spectra, and (2) the inference of long-time dynamics with the aid of the transfer tensor method (TTM).{\cite{cerrillo2014prl}}

We first focus on the linear absorption spectra. We assume the dipole operator is given by $\hat{\mu} = \hat{\sigma}_{x}$. The dipole-dipole correlation function is defined as 
\begin{equation}
C(t) = {\rm{Tr}}_{\rm{tot}}\left\{ \hat{\mu} \, e^{-i \hat{H}_{\rm{tot}} t} \,\left( \hat{\mu} 
\hat{\rho}_{\rm{s}}(0) \otimes \hat{\rho}_{\rm{b}} \right) e^{i \hat{H}_{\rm{tot}} t} \right\},
\label{eq.corr_1st}
\end{equation}
where $\hat{\rho}_{\rm{s}}(0) = |2\rangle \langle 2|$, and 
$\hat{\rho}_{\rm{b}}$ is the thermal equilibrium density matrix of the bath. The linear absorption spectrum is obtained via a Fourier transformation 
\begin{equation}
C(\omega) = {\rm{Re}}\int_{0}^{\infty} {\rm{d}}t \, C(t) \, e^{-i \omega t}.
\end{equation}
To evaluate Eq.~{\eqref{eq.corr_1st}} within the NMQSD framework, we employ the following decomposition {\cite{chen2022jcp}}
\begin{equation}
\hat{\mu} \hat{\rho}_{\rm{s}}(0) = \sum_{\eta} \eta |\psi_0(\eta)\rangle \langle \psi_0(\eta)|,
\label{eq.off_diag_sum}
\end{equation}
where $|\psi_0(\eta)\rangle = (|1\rangle + \eta |2\rangle)/\sqrt{2}$ with $\eta \in \{ \pm 1,  \pm i \}$. This decomposition is necessary because NMQSD is formulated for pure initial states and cannot directly handle coherence terms. Using Eqs.{\eqref{eq.reduced_density}} and {\eqref{eq.operator_evolve}}, we obtain 
\begin{equation}
C(t) = \sum_{\eta} \frac{\eta}{2} \, \mathbb{E} \langle \psi_0(\eta)| 
\hat{U}^{\dagger}(t, \eta) \hat{U}(t, \eta) |  \psi_0(\eta) \rangle,
\end{equation}
where $\hat{U}(t,\eta)$ is the precomputed time evolution operator for the initial state $|\psi_0\rangle = |  \psi_0(\eta) \rangle$.

\begin{figure}
\centering
\includegraphics[width=\figratio\textwidth]{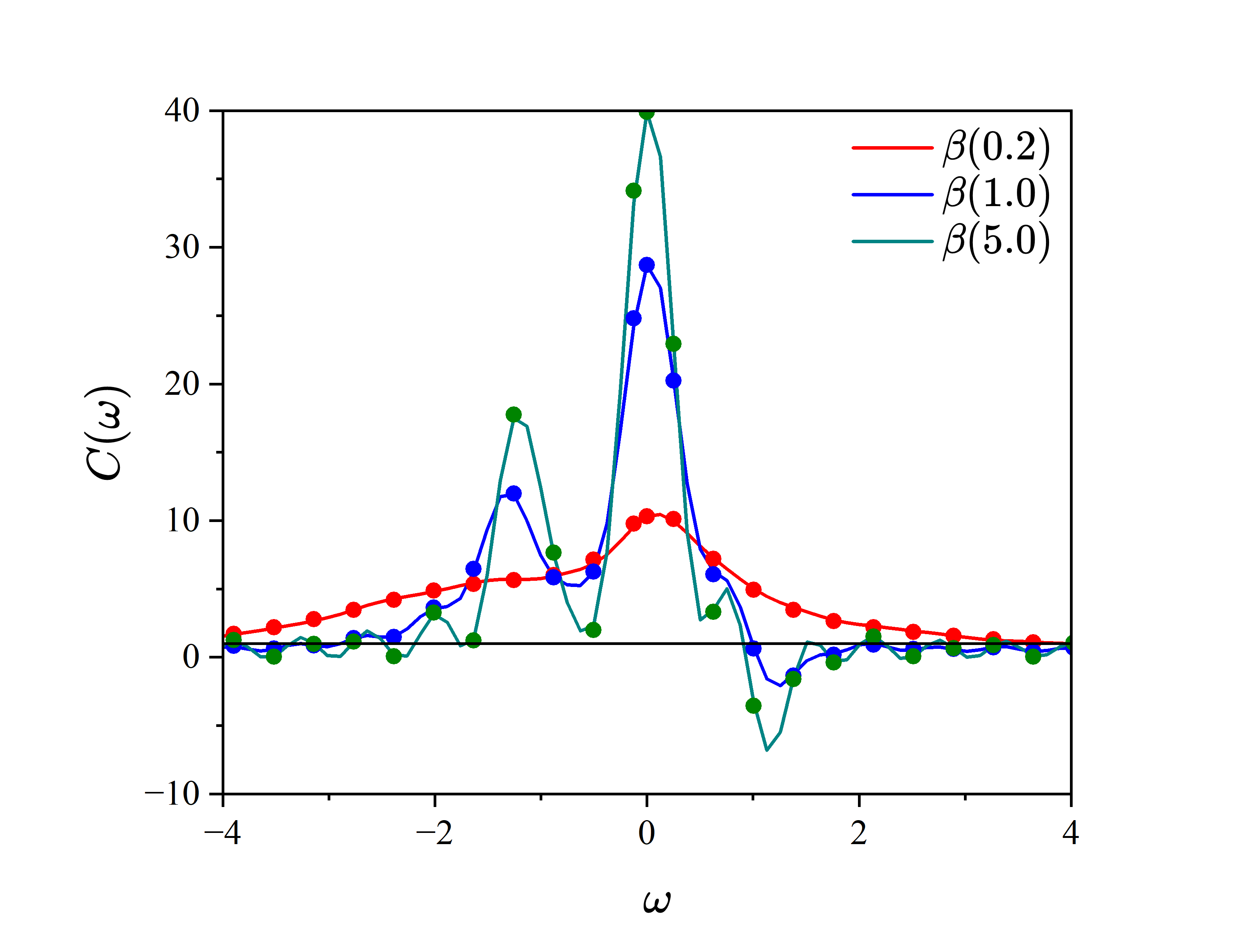}
\caption{Linear absorption spectra $C(\omega)$ obtained from the constructed time evolution operator (solid curve) and the HEOM method (dot) at $\beta = 0.2$ (red), $1$ (blue), and $5$ (green), respectively.}
\label{fig.resp_drude}
\end{figure} 

In Fig.~{\ref{fig.resp_drude}}, we show the linear absorption spectra obtained from the constructed time evolution operator. The ensemble average over the noise $z_t$ was conducted over the entire dataset, including both training and validation samples. The reference result was computed by directly propagating Eq.~{\eqref{eq.corr_1st}} using the HEOM method. To avoid large anomalies near the boundary, the propagation time was limited to $t \in (0, \, 0.8 t_{\rm{max}})$. Consequently, the Fourier transform from this finite time window results in some minor negative values around $\omega \simeq 1$. Despite this numerical artifact, the spectroscopic profile from the constructed time evolution operator remains quantitatively consistent with the HEOM reference. 

Due to the memory integral inherent in NMQSD, the composition property does not hold, i.e., $|\tilde{\psi}_{2t}\rangle \ne \hat{U}(t) \hat{U}(t) |\psi_0\rangle$. Therefore, a trianed model cannot be simply extended beyond its original time window $t \in (0, t_{\rm{max}})$. However, the primary quantity of physical interest is generally the reduced density matrix $\hat{\rho}_{\rm{s}}(t)$, not the stochastic wavefunction $|\tilde{\psi}_{t}\rangle$. To compute the time evolution for $t > t_{\rm{max}}$, we employ the TTM.{\cite{cerrillo2014prl, strachan2024jcp}} In this approach, a set of transfer tensors is constructed systematically from the short-time evolution operator $(t < t_{\rm{max}})$. These tensors are then used to infer the long-time dynamics via a convolutional equation. The details of this calculation are provided in Appendix {\ref{sec.app.qme}}. In Fig.~{\ref{fig.pop_drude}}, we show the population dynamics $\Delta(t) = \langle 1|\hat{\rho}_{\rm{s}}(t) |1\rangle - 
\langle 2|\hat{\rho}_{\rm{s}}(t) |2\rangle$ extended to $t = 40$ $(4 t_{\rm{max}})$ for various values of $\beta$ using the TTM, with the initial state $\hat{\rho}_{\rm{s}}(0) = |1\rangle \langle 1|$. The reference result was obtained from the HEOM method, using the same hierarchy settings as for HOPS. The consistency of the results, even for $t > t_{\rm{max}}$, demonstrates the accuracy of our model and highlights the broad utility of the precomputed time evolution operators.

\begin{figure}
\centering
\includegraphics[width=\figratio\textwidth]{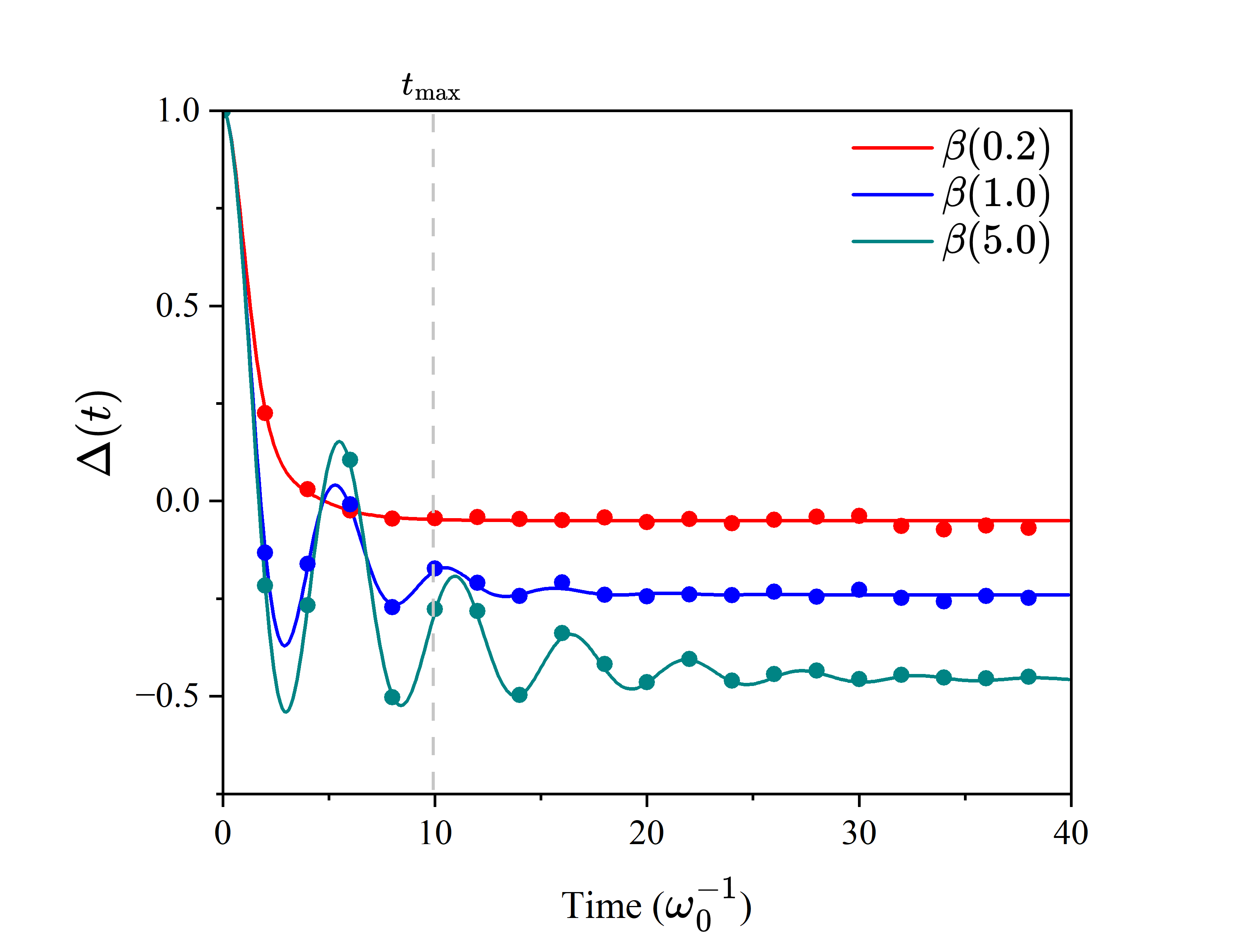}
\caption{Population dynamics $\Delta(t) = \langle 1|\hat{\rho}_{\rm{s}}(t) |1\rangle - 
\langle 2|\hat{\rho}_{\rm{s}}(t) |2\rangle$ obtained from our model with the aid of TTM (solid curve) and HEOM reference (dot) at $\beta = 0.2$ (red), $1$ (blue), and $5$ (green), respectively.
Result from our model coincides with reference results even beyond training time window $t_{\rm{max}}$ (highlighted as gray dashed line). }
\label{fig.pop_drude}
\end{figure} 

\section{Conclusion}
\label{sec.conclusion}

In summary, we presented a novel application of neural networks for solving the NMQSD equation by introducing a new learning algorithm designed for operator construction. This algorithm defines a time-evolution operator for the underlying stochastic process and develops a universal method to construct it from time-dependent wavefunctions using a neural network. Furthermore, we extend this approach to construct the operator ansatz for the functional derivative within the NMQSD framework. Once constructed, these operators can be directly utilized to solve the NMQSD equation based on their formal definitions. Crucially, they can also be integrated into other computational tasks independently of the original neural network. This provides significantly greater flexibility compared to conventional machine learning algorithms, which often focus solely on approximating the wavefunction or expectation values, an approach with limited applicability. For the numerical implementation, we developed a U-Net inspired architecture composed of multiple parallel blocks, each containing a different number of Fourier layers. This design allows for a more efficient representation of highly mixed quantum correlations across multiple scales. We benchmarked our model on a typical spin-boson system, demonstrating its accuracy over broad parameter regimes.

This work establishes a foundation for further exploration of neural network based operator construction. One promising application is the generation of transfer tensors for non-Markovian quantum master equations.{\cite{lyu2023jctc, chen2020pra, sayer2024jcp}} For time-independent Hamiltonians, a transfer tensor built from short-time data can predict long-time dynamics; our algorithm could serve as a universal and numerically efficient method for this construction. Another key direction for improvement involves seeking more efficient operator representations. While this work uses explicit matrices, which scale exponentially with system size, future efforts could target formats like a weighted set of Pauli strings. Developing hybrid neural network and quantum computing algorithms to reconstruct such operators from observable data represents an important future research direction.

\section*{Acknowledgments}
J.Z. and L.P.C. acknowledge support from the National Natural Science Foundation of China (Grant No. 22473101). 
C.L.B.-R. gratefully acknowledges financial support from the Royal Society of Chemistry and the European Union’s Horizon Europe Research and Innovation program under the Marie Skłodowska-Curie Grant Agreement No. 101065295-RDMFTforbosons.

\appendix

\section{Hierarchy of pure state}
\label{sec.app.hops}

The HOPS method avoids the explicit processing of functional derivatives by introducing a set of hierarchical elements, analogous to the density matrix-based HEOM method. This converts the original integro-differential NMQSD into a closed set of time-local differential equations defined within these hierarchies. 

Let us assume a linear decomposition of the bath correlation function,
\begin{equation}
\alpha(t) = \sum_{k}^{K} c_{k} e^{-\gamma_{k} |t|},
\label{eq.hops.linear_sum}
\end{equation}
the non-linear form of HOPS can be derived using the Girsanov transformation 
\begin{equation}
\begin{split}
\partial_t | {\tilde{\psi}}_{\bm{h}}(t) \rangle &= \left( -i \hat{H}_{\rm{s}} 
- \sum_{k}^{K} h_{k} \gamma_{k} + \tilde{z}_{t} \hat{V} \right) | {\tilde{\psi}}_{\bm{h}}(t)\rangle \\
&+ \sum_{k}^{K} c_{k} \hat{V} | {\tilde{\psi}}_{\bm{h}_{-k}}(t) \rangle 
- \left(\hat{V}^{\dagger} - \langle \hat{V}^{\dagger}\rangle_t  \right)\sum_{k}^{K} 
| {\tilde{\psi}}_{\bm{h}_{+k}}(t) \rangle. 
\end{split}
\label{eq.hops}
\end{equation}
Here, the index vector ${\bm{h}} = \{ h_1, h_2, \cdots, h_{K} \}$ consists of non-negative integers, and ${\bm{h}}_{\pm k}$ denotes $\{ \cdots, h_{k} \pm 1, \cdots \}$. Analogous to HEOM, the $0$-th hierarchical element corresponds to the state, $| \tilde{\psi}_t \rangle$, from Eq.~{\eqref{eq.nmqsd_nonlinear}}. All other elements are used to account for non-perturbative and non-Markovian effects.  

In the simulations, we employed a Pad\'{e} decomposition for Eq.~{\eqref{eq.hops.linear_sum}}.{\cite{ikeda2020jcp, hu2011jcp}} The hierarchy was truncated at $K = 5 \sim 10$, with $\sum_{k} h_{k} \le 20$, based on the parameters $\gamma$ and $\beta$. The HOPS equations were propagated using a low-storage fourth-order Runge-Kutta solver with a time step of $\delta_t = 0.01$. The dataset was generated using the chosen initial state and noise $z_t$ as discussed in Section {\ref{sec.result}}.

\section{Transfer tensor method from the trained model}
\label{sec.app.qme}

This section illustrates how to infer the long-time dynamics for $t > t_{\rm{max}}$ using the time evolution operator of the NMQSD, constructed from our model. We adopt the TTM to describe the time evolution of $\hat{\rho}_{\rm{s}}(t)$ under a time-independent Hamiltonian. {\cite{cerrillo2014prl, strachan2024jcp}} We begin by defining a discrete time grid, $t_{n} = n \delta_t$, and a corresponding dynamical map
\begin{equation}
\hat{\rho}_{\rm{s}}(t_n) = \mathcal{E}_{n} \hat{\rho}_{\rm{s}}(0),
\end{equation}
where the super-operator $\mathcal{E}_{n}$ has dimensions of $N_{\rm{s}}^{2} \times N_{\rm{s}}^{2}$, with $N_{\rm{s}}$ being the dimension of the system. Next, we introduce a set of transfer tensors, $\mathcal{T}_{k}$, which share the same dimensions as the dynamical map. These tensors define the evolution through a recursive relation  
\begin{equation}
\hat{\rho}_{\rm{s}}(t_n)  = \sum_{k=1}^{K} \mathcal{T}_{k} \hat{\rho}_{\rm{s}}(t_{n-k}),
\label{eq.nmqme_tt}
\end{equation}
where $K$ is a chosen cutoff level such that $\mathcal{T}_{k} = 0$ for $k > K$. The transfer tensors can be obtained recursively from the dynamical map using the relation
\begin{equation}
\mathcal{T}_{n} = \mathcal{E}_{n} - \sum_{k=1}^{n-1} \mathcal{T}_{n - k} \mathcal{E}_{k} .
\label{eq.nmqme_dynmap_tt}
\end{equation}

For the spin-boson model, the elements of the dynamical map are computed from the time evolution operator in Eq.~{\eqref{eq.reduced_density}} as follows
\begin{equation}
\mathcal{E}_{n}\bigr|_{kj, 11} = \mathbb{E}\left[ \langle k | \hat{U}(t_n, 1) |1\rangle 
\langle 1| \hat{U}^{\dagger}(t_n, 1) | j \rangle \right].
\end{equation}
\begin{equation}
\mathcal{E}_{n}\bigr|_{kj, 22} = \mathbb{E}\left[ \langle k | \hat{U}(t_n,1) |2\rangle 
\langle 2| \hat{U}^{\dagger}(t_n, 1) | j \rangle \right],
\end{equation}
with the abbreviation $\hat{U}(t_n, k) \equiv \hat{U}(t_n, {\bm{z}}_{[0:n]}, |k\rangle)$.
For off-diagonal elements, we employ the decomposition,
\begin{equation}
|1\rangle \langle 2| = \sum_{\eta \in \{ \pm 1, \pm i\}} \frac{\eta}{2} |\psi_0(\eta)\rangle \langle \psi_0(\eta)|,
\end{equation}
where $|\psi_0(\eta)\rangle = (|1\rangle + \eta |2\rangle)/\sqrt{2}$. The corresponding dynamical map element is 
\begin{equation}
\mathcal{E}_{n}\bigr|_{kj, 12} = \sum_{\eta} \frac{\eta}{2} ~ \mathbb{E}\left[ 
\langle k | \hat{U}(t_n, \eta) |\psi(\eta)\rangle \langle \psi(\eta) | 
\hat{U}^{\dagger}(t_n, \eta) | j \rangle \right],
\end{equation}
with the 
abbreviation  $\hat{U}(t_n, \eta) \equiv \hat{U}(t_n, {\bm{z}}_{[0:n]}, |\psi_0(\eta)\rangle)$. The transfer tensors are constructed from Eq.~{\eqref{eq.nmqme_dynmap_tt}} using a time step of $\delta_t = 0.01$ and a cutoff of $K=500$ (corresponding to $K\delta_t = 0.5 t_{\rm{max}}$). Once constructed, the dynamics to arbitrarily long times can be efficiently inferred by iterating Eq.~{\eqref{eq.nmqme_tt}}.

\bibliography{ref_hops_abbrev}

\end{document}